# Monomorphous decomposition method for phase retrieval and phase-contrast tomography

T.E. Gureyev[1,2,3] and Ya.I. Nesterets[1,3]

[1] *Commonwealth Scientific and Industrial Research Organisation, Clayton, VIC 3168, Australia;*
[2] *Monash University, Clayton, VIC 3800, Australia;*
[3] *University of New England, Armidale, NSW 2358, Australia.*

9 April 2015

**Abstract**

We show that an arbitrary spatial distribution of complex refractive index inside an object can be exactly represented as a sum of two "monomorphous" complex distributions, i.e. the distributions with the ratios of the real part to the imaginary part being constant throughout the object. A priori knowledge of constituent materials can be used to estimate the global lower and upper boundaries for this ratio. This approach can be viewed as an extension of the successful phase-retrieval method, based on the Transport of Intensity equation, that was previously developed for monomorphous (homogeneous) objects, such as e.g. objects consisting of a single material. We demonstrate that the monomorphous decomposition can lead to more stable methods for phase retrieval using the Transport of Intensity Equation. Such methods may find application in quantitative in-line phase-contrast imaging and phase-contrast tomography.

*1. Introduction*

The use of the Transport of Intensity equation (TIE) [1, 2] for solving the problem of optical phase retrieval, i.e. the problem of reconstruction of the phase distribution of the complex amplitude of a free-propagating optical beam from one or more measurements of its intensity distribution in the plane(s) orthogonal to the optic axis in the Fresnel region, has been extensively investigated since the publication of a seminal paper by Teague [2] in 1992. The approach was successfully applied in infrared adaptive optics [3], electron microscopy [4, 5] and later in X-ray imaging [6-23], visible light microscopy [24, 25] and elsewhere. The success of the method led to a large number of studies which reported various implementations of the TIE-based phase retrieval and the validity limits for the method. In particular, it has been shown [12] that in order to achieve quantitatively accurate phase retrieval, the propagation-induced contrast (i.e. the difference between the image distributions in the object and image planes) must be weak. This condition typically leads to low signal-to-noise ratios and, consequently, to poor numerical stability in the associated phase retrieval which affects primarily the low-spatial-frequency components of the reconstructed phase distributions. Here we propose a method that can potentially alleviate this instability with the help of generic *a priori* information about the sample.

The next section of the paper contains an overview of the so-called "homogeneous" or "monomorphous" version of the TIE [8] and related approaches. Section 3 describes a monomorphous decomposition of a generic complex refractive index and complex wave amplitudes. Section 4 presents several versions of monomorphous representation of the TIE and discusses their possible applications in in-line imaging, phase retrieval and phase-contrast tomography. In Section 5 we test some of the methods developed in Section 4 using a numerical phantom. Section 6 contains a brief summary.

*2. Monomorphous TIE and the problem of stability of in-line phase retrieval*

Most methods for phase retrieval using the TIE require multiple X-ray projection images to be acquired under appropriately varied conditions, in order to reconstruct the phase and intensity distributions in the object plane. Suitable projection images can be collected at two or more different sample-to-detector distances [2] or at different X-ray energies [26]. Known exceptions to this rule, where a single image per view angle is sufficient for an exact reconstruction, are represented by the following three cases.

(1) Conventional (or "contact") transmission X-ray imaging and CT, which can be viewed as a limit case of in-line imaging, in which the sample-to-detector distance is negligibly small. Here X-ray refraction effects do not contribute to the registered images and as a result only the (projection of the) imaginary part $\beta(\mathbf{r})$ of the complex refractive index $n(\mathbf{r}) = 1 - \delta(\mathbf{r}) + i\beta(\mathbf{r})$, $\mathbf{r} = (x, y, z)$, which is responsible for differential absorption of X-ray in the sample, is reconstructed.

(2) The opposite case is represented by the so-called pure-phase objects which exhibit negligible absorption at the X-ray energies used in the experiment. Here only the (projection of the) real decrement $\delta(\mathbf{r})$ of the complex refractive index contributes to the image contrast and can be reconstructed in in-line imaging experiments.

(3) Finally, there is a class of samples characterized by a fixed proportionality relationship between the real and imaginary parts of the refractive index decrement [8]:

$$\delta(\mathbf{r}) / \beta(\mathbf{r}) = \gamma, \qquad (1)$$

where $\gamma$ does not depend on $\mathbf{r}$. Obviously, this relationship reduces the number of unknown functions from two to just one (assuming that $\gamma$ is known a priori) and therefore a single projection is sufficient for the reconstruction of both intensity and the phase. Such objects are sometimes also called "monomorphous" [27]. They include, for example, "homogeneous" samples which consist predominantly of a single material whose density may vary spatially. In fact, the above classes (1) and (2) can be viewed as special cases of class (3) with $\gamma = 0$ and $\gamma = \infty$, respectively.

We consider here the case of a plane monochromatic incident wave of unit intensity propagating along the optical axis $z$. The object that is being imaged, is located in the half-space $z < 0$ and is characterized by the distribution of its complex refractive index. The value of eq.(1) for TIE-based phased retrieval is in the following relationship (valid under the projection approximation) between the phase, $\varphi_0(x, y) = -k\int_{-\infty}^{0} \delta(x, y, z)dz$, and intensity, $I_0(x, y) = \exp[-2k\int_{-\infty}^{0} \beta(x, y, z)dz]$, distributions in the object plane:

$$\varphi_0(x, y) = (\gamma/2)\ln I_0(x, y). \qquad (2)$$

Therefore, for the purpose of phase retrieval it does not matter if eq.(1) holds, as long as the (weaker) equation (2) is satisfied. Moreover, it turns out that in order to derive the "monomorphous" version of the general TIE, it is sufficient to require the constant proportionality only between the gradients of the logarithm of intensity of the transmitted wave and the logarithm of its phase across the object plane:

$$\nabla\varphi_0(x, y) = (\gamma/2)\nabla I_0(x, y)/I_0(x, y), \qquad (3)$$

where $\nabla \equiv (\partial_x, \partial_y)$ is the 2D transverse gradient operator. Obviously, eq.(3) is a weaker requirement compared to that of the proportionality of the logarithm of intensity and the phase themselves, as in eq.(2). Substituting eq.(3) into the generic TIE [1, 2], one arrives at the following "monomorphous" form of the TIE [8]:

$$I_R(x, y) = I_0(x, y) - \gamma R/(2k)\nabla^2 I_0(x, y), \qquad (4)$$

where $I_R(x, y)$ is the intensity distribution in the detector (image) plane $z = R$.

Given the inherent sensitivity of the phase-retrieval methods based on the TIE to noise in the input images, the technique proposed in [8] on the basis of eq.(4) was an important development as it allowed a stable and quantitatively accurate recovery of the phase from a single in-line image containing realistic amounts (several percent) of noise. To date, this "monomorphous TIE" method appears to be by far the most successful one in X-ray imaging applications of the TIE. The remarkable stability of the method has been explained by the fact that it optimally combines the sensitivity of the phase contrast to high-spatial-frequency components of the transmitted complex amplitude, provided by the second term on the right-hand side (r.h.s.) of eq.(4), with the complementary sensitivity (provided by the first term on the r.h.s. of eq.(4)) of the absorption contrast to the low-frequency components. Mathematically, when $\gamma > 0$, this equation corresponds to a strictly positive partial differential operator whose spectrum is separated from zero and, therefore, it does not have zeros in the corresponding contrast transfer function at low spatial frequencies. However, this useful property can only be attained for a special class of objects (transmitted complex amplitudes), namely those satisfying eq.(3). Samples consisting of a single material obviously possess this property [8], as well as samples consisting of light chemical elements with $Z < 10$, if the X-ray energy of the incident radiation is approximately between 60 and 500 keV [28]. Due to the proportionality of the attenuation and phase shifts generated by such samples, the phase can be retrieved from a single defocused image [8], which is of course an extremely useful property as it allows one to avoid experimental problems related to image co-registration due to possible instabilities of the incident beam, optical elements and/or the sample, as well as significantly simplify the data acquisition compared to phase-retrieval methods requiring the acquisition of multiple images. The applicability of this method to monomorphous samples only is the main limitation of the method.

As a further natural extension of the "monomorphous" condition represented by eq.(3), we would like to mention the following theorem proven in [29]:

for an arbitrary pair of suitably well-behaved functions $(I_0, \varphi_0)$ in a domain $\Omega$ in a 2D plane $(x, y)$, there exists a function $\psi$ such that $\nabla \psi_0(x, y) = I_0(x, y) \nabla \varphi_0(x, y)$, if and only if

$$\nabla I_0(x,y) \times \nabla \varphi_0(x,y) \equiv 0, \tag{5}$$

(where "×" denotes the vector product), i.e. if the vector fields $\nabla \varphi_0$ and $\nabla I_0$ are parallel to each other everywhere in $\Omega$ (here a vector of zero length is considered parallel to any other vector). Note that eq.(3) means that the vectors $\nabla \varphi_0$ and $\nabla I_0$ are parallel everywhere, and the ratio of their lengths is equal exactly to $(\gamma/2)/I_0(x,y)$ at each point. Therefore, eq. (5) is indeed a direct generalization of eq.(3). The above equivalence of eq.(5) and the existence of gradient function $\psi$ such that $\nabla \psi_0(x,y) = I_0(x,y) \nabla \varphi_0(x,y)$ means that eq.(5) is a sufficient condition for the well-known method for solution of the TIE originally proposed by Teague [2] and later developed in [30] and used in many other publications. Note however that, unlike the phase retrieval using the homogeneous TIE eq.(4), Teague's method, being based on the generic TIE, requires at least two different intensity images acquired e.g. at different defocus distances. As a consequence, Teague's method does not deliver any extra stability to the solution of the TIE compared to other, more generic methods. It does lead, however, to a form of "single-step" phase-contrast CT reconstruction formula [31] that generalises the result originally obtained by Bronnikov [32, 33] and later extended by others [34, 35].

A number of different methods for TIE-based phase retrieval have been proposed and tested for generic (non-monomorphous) objects [9-20]. These methods usually require more than one image collected either at different defocus distances [2] or at the same distance but at different radiation wavelengths [12]. While being formally mathematically well-posed [6], these methods suffer from the generic low-frequency instability inherent to phase retrieval using the TIE. As mentioned above, this instability is tightly related to the requirement for the propagation contrast to be low in order for the TIE approximation to be valid. Although quantitatively accurate phase retrieval from multiple defocused images of a generic object has been demonstrated in the visible light region [24, 25], as far as we are aware this success has never been reproduced convincingly with X-rays, despite a number of attempts. It appears that one of the main difficulties in performing accurate TIE-based phase retrieval from multiple defocused X-ray images is in the variation of the incident illumination, which tends to be more pronounced for X-ray sources compared to high-quality visible light sources.

While the change in the incident intensity can often be at least partially compensated by using appropriate "flat field" images (collected at the same defocus distances but without the sample), the change in the phase distribution of the incident illumination generally cannot be corrected for, except, perhaps, for the lowest tilt and defocus aberrations that can be detected and corrected in software by comparing the positions of image boundaries. The other aberrations of the illuminating beam usually end up as artefacts in the reconstructed phase, which often overwhelm the true signal from the sample.

The above situation is fairly standard for reconstructive imaging under low signal-to-noise conditions. One of the most powerful tools for dealing with this type of problems is the use of a priori information. Obviously, in order to maximize the usability of a phase-retrieval method one would generally want to minimize the amount of a priori information required for successful performance of the method, and, whenever possible, use only generic information, such as e.g. the positivity of the real and imaginary parts of the complex refractive index. Given the success of the monomorphous TIE method, it appears useful to try to extend its positive traits, namely the use of absorption contrast for regularizing phase retrieval at low spatial frequencies, to generic samples. Even though for generic samples one cannot assume that the ratio of the real to imaginary part of the refractive index is constant throughout the sample, it should be possible in most cases to estimate the upper and lower limits of this ratio, e.g. from a priori knowledge of the expected material constituents of the sample. By constraining this ratio one should be able to eliminate at least some of the phase artefacts, thus improving the stability of the phase retrieval. This constitutes the basic idea of the method presented below.

## 3. Monomorphous decomposition of complex refractive index and complex wave amplitude

The interaction of an object with an incident monochromatic X-ray beam is determined by a 3D distribution of its complex refractive index:

$$n(\mathbf{r};\lambda) = 1 - \delta(\mathbf{r};\lambda) + i\beta(\mathbf{r};\lambda), \tag{6}$$

where $\mathbf{r} = (x, y, z)$ is the Cartesian spatial coordinate and $\lambda$ is the X-ray wavelength. We have adopted the definition according to which an object is called "monomorphous" if the ratio $\gamma(\lambda) \equiv \delta(\mathbf{r}; \lambda) / \beta(\mathbf{r}; \lambda)$ is independent from $\mathbf{r}$ throughout the sample. We will omit below the wavelength argument $\lambda$ for brevity.

Let us check now that an arbitrary complex refractive index distribution can be represented as a sum of two monomorphous ones, or more precisely that for any pair of constants $\gamma_1$ and $\gamma_2$, $\gamma_1 \neq \gamma_2$, there exist such real-valued functions $\beta_1(\mathbf{r})$ and $\beta_2(\mathbf{r})$ that

$$-\delta(\mathbf{r}) + i\beta(\mathbf{r}) \equiv \beta_1(\mathbf{r})(-\gamma_1 + i) + \beta_2(\mathbf{r})(-\gamma_2 + i). \tag{7}$$

Indeed, it is easy to verify that eq.(7) is satisfied, provided that

$$\begin{cases} \beta_1(\mathbf{r}) = [\delta(\mathbf{r}) - \gamma_2 \beta(\mathbf{r})]/(\gamma_1 - \gamma_2) \\ \beta_2(\mathbf{r}) = [\delta(\mathbf{r}) - \gamma_1 \beta(\mathbf{r})]/(\gamma_2 - \gamma_1) \end{cases}. \tag{8}$$

As we see, eq.(7) has a unique solution for any pair of constants $\gamma_1$ and $\gamma_2$, such that $\gamma_1 \neq \gamma_2$. However, normally for X-rays $\beta(\mathbf{r}) > 0$ and $\delta(\mathbf{r}) > 0$, and therefore it is natural to demand that both $\beta_1(\mathbf{r})$ and $\beta_2(\mathbf{r})$ are positive everywhere (assuming that both $\gamma_1$ and $\gamma_2$ are positive as well). Therefore, if for example $0 < \gamma_1 < \gamma_2$, it is easy to verify that the requirement for $\beta_1(\mathbf{r})$ and $\beta_2(\mathbf{r})$ to be positive leads to the following condition:

$$\gamma_1 \leq \delta(\mathbf{r}) / \beta(\mathbf{r}) \leq \gamma_2 \text{ for all } \mathbf{r} \text{ inside the sample.} \tag{9}$$

Condition eq.(9) implies a strategy where a monomorphous decomposition of an unknown object should have the first monomorphous component with the minimal $\delta(\mathbf{r})/\beta(\mathbf{r})$ ratio for all materials possibly present in the sample, while the second component should have the maximal $\delta(\mathbf{r})/\beta(\mathbf{r})$ ratio. In this context, it is already clear that the two monomorphous components establish a priori "envelope" for the reconstructed values of the $\delta(\mathbf{r})/\beta(\mathbf{r})$ ratio, thus preventing large erroneous phase oscillations which otherwise might have appeared due to inconsistencies in the measured image intensities.

Let the monochromatic plane incident X-ray wave $I_{in}^{1/2}\exp(ikz)$ with intensity $I_{in}$ and wavevector $k = 2\pi/\lambda$ propagate along the optic axis $z$. Given the monomorphous decomposition eq.(7) of the sample, we can represent the transmitted intensity in the object plane $z = 0$ as

$$I_0(x, y) = I_{in} Q_1(x, y) Q_2(x, y), \tag{10}$$

where

$$Q_j(x, y) = \exp[-2k\int \beta_j(x, y, z)dz], \quad j = 1, 2, \tag{11}$$

are the transmittances corresponding to the two monomorphous components. The corresponding transmitted phase in the object plane is then

$$\varphi_0(x, y) = \varphi_1(x, y) + \varphi_2(x, y), \quad . \tag{12}$$

where

$$\varphi_j(x, y) = -k \int \delta_j(x, y, z)dz) = (\gamma_j / 2) \ln Q_j(x, y), \quad j = 1, 2. \tag{13}$$

Note that the monomorphous decomposition, eqs.(10)-(13), of a transmitted complex amplitude $U_0(x, y) \equiv I_0^{1/2}(x, y)\exp[i\varphi_0(x, y)]$ can in principle be performed without a reference to the monomorphous decomposition eq.(7) of the object, using instead some "abstract" transmission functions $Q_1$ and $Q_2$. Indeed, for any given pair of functions $I_0(x, y)$ and $\varphi_0(x, y)$, and any pair of constants $0 < \gamma_1 < \gamma_2$, such that

$$\gamma_1 \leq 2\varphi_0(x, y) / b_0(x, y) \leq \gamma_2 \text{ for all } (x, y), \tag{14}$$

where $b_0(x, y) \equiv \ln[I_0(x, y)/I_{in}]$, there exists a unique pair of functions $Q_1(x, y)$ and $Q_2(x, y)$, such that equations (10) and (12) hold with $\varphi_j(x, y) = (\gamma_j / 2)\ln Q_j(x, y), \ j = 1, 2$. It is straightforward to verify that the required functions are given by:

$$\begin{cases} \ln Q_1(x, y) = [2\varphi_0(x, y) - \gamma_2 b_0(x, y)]/(\gamma_1 - \gamma_2) \\ \ln Q_2(x, y) = [2\varphi_0(x, y) - \gamma_1 b_0(x, y)]/(\gamma_2 - \gamma_1) \end{cases}. \tag{15}$$

Condition eq.(14) ensures that the functions $Q_j(x, y)$ satisfy the inequalities

$$I_0(x, y)/I_{in} \leq Q_j(x, y) \leq 1 \text{ for all } (x, y), j = 1, 2. \tag{16}$$

The limit cases: $Q_1(x, y) = 1$, $Q_2(x, y) = I_0(x, y)/I_{in}$, and $Q_1(x, y) = I_0(x, y)/I_{in}$, $Q_2(x, y) = 1$, correspond to monomorphous cases with $\varphi_0(x, y) = (\gamma_2/2)b_0(x, y)$ and $\varphi_0(x, y) = (\gamma_1/2)b_0(x, y)$, respectively.

*4. Transport of Intensity equation in monomorphous representation*

The TIE expresses intensity distribution in a plane $z = R$ downstream from the object plane $z = 0$, as a function of the intensity and phase distributions in the object plane [1, 2]:

$$I_R(x, y) = I_0(x, y) - (R/k)\nabla \cdot [I_0(x, y)\nabla\varphi_0(x, y)]. \qquad (17)$$

Substituting the monomorphous representation eqs.(10)-(13) for the intensity and phase in the object plane into eq.(17) and omitting all function arguments for brevity, we obtain

$$I_R = [1 - R\gamma_1/(2k)\nabla^2]I_0 - R(\gamma_2 - \gamma_1)/(2k)I_{in}\nabla \cdot [Q_1\nabla Q_2]. \qquad (18)$$

This equation should be considered together with eq.(10), $I_0 = I_{in}Q_1Q_2$. If we express $Q_1 = I_0/(I_{in}Q_2)$ from eq.(10) and substitute it into eq.(18), the latter equation becomes

$$I_R = [1 - R\gamma_1/(2k)\nabla^2]I_0 - R(\gamma_2 - \gamma_1)/(\gamma_2 k)\nabla \cdot [I_0\nabla\varphi_2]. \qquad (19)$$

This equation can be solved for the unknown phase $\varphi_2$ if intensity distributions in the object and image planes, $I_R$ and $I_0$, are known. Subsequently, $Q_2$ can be easily calculated as $Q_2 = \exp(2\varphi_2/\gamma_2)$, then $Q_1$ can be obtained as $Q_1 = I_0/(I_{in}Q_2)$. This gives us $\varphi_1 = (\gamma_1 / 2) \ln Q_1$ and finally $\varphi_0 = \varphi_1 + \varphi_2$. Unfortunately, eq.(19) is even worse in terms of numerical stability than the original TIE, eq.(17). In order to emulate the favourable stability properties of the monomorphous TIE we shall re-arrange eq.(18) as follows:

$$2I_R - [1 - (R\gamma_1 / k)\nabla^2]I_0 = I_{in}Q_1Q_2 - [R(\gamma_2 - \gamma_1)/k]I_{in}\nabla \cdot [Q_1 \nabla Q_2]. \qquad (20)$$

The last equation does possess the desired mathematical stability property with respect to the unknown function $Q_2$ (if $Q_1$ is known). Indeed, it is easy to see that: (a) the first term on the r.h.s. of eq.(20) represents a multiplication of function $Q_2$ by a function $I_{in}Q_1$ which is positive everywhere; (b) the second term on the r.h.s. of eq.(20) represents a non-negative partial-differential operator applied to $Q_2$. Hence, the whole of the r.h.s. of eq.(20) represents a strictly positive operator applied to $Q_2$ (i.e. the spectrum of this operator is separated from zero). Therefore, this operator is invertible and the norm of its inverse (determined by the inverse of the lower bound of the spectrum of the direct operator) is finite, i.e. eq.(20) is mathematically well-posed and stable.

We can solve eq.(20) in combination with eq.(10) iteratively. We can take, for example, $Q_1^{(0)} \equiv 1$ as an initial iteration (a similar technique can be applied with $Q_1^{(0)} \equiv I_0$, and with other choices). This transforms eq.(20) into a conventional monomorphous TIE which can be explicitly solved:

$$Q_2^{(0)} = I_{in}^{-1}[1 - R\gamma_2 /(2k)\nabla^2]^{-1} I_R. \qquad (21)$$

In other words, the zero-order approximate solution is given here by the monomorphous distribution with $2\varphi_0^{(0)}(x, y) = \gamma_2 b_0(x, y)$. Subsequent iterations are performed by evaluating $Q_1^{(n)} = I_0 /(I_{in}Q_2^{(n-1)})$, substituting this into eq.(20) and solving the resultant equation:

$$Q_2^{(n)} = [I_{in}Q_1^{(n)} - Rk^{-1}(\gamma_2 - \gamma_1)I_{in}\nabla \cdot (Q_1^{(n)}\nabla)]^{-1}[2I_R - (1 - Rk^{-1}\gamma_1\nabla^2)I_0]. \tag{22}$$

The latter solution can be implemented numerically e.g. using the Full Multigrid method [36]. Equation (22) has stability properties similar to those of the monomorphous TIE due to the fact that $Q_1^{(n)}(x, y) \geq const > 0$ at any $n$.

The issue of convergence of the iterative process is not obvious, and we can only state that we have observed proper convergence in the numerical examples that we have considered so far in the absence of noise. If the input data contains some noise, then the iterative process displays the usual semi-convergent behaviour, i.e. it converges for several iterations before beginning to diverge. In this context it becomes important to find a reliable "stopping criterion" in order to prevent the process from diverging. For example, one can stop the iterations when the difference between successive iterations becomes smaller than the noise level. Assuming Poisson noise, this leads to the following criterion:

$$\| I_0^{1/2}[1 - Q_2^{(n)}/Q_2^{(n-1)}] \|_2 > \sigma_{bckg}, \tag{23}$$

where $\sigma_{bckg}$ is the standard deviation of the Poisson distribution in a sample-free area of the image (background) and $\| \|_2$ denotes the normalized Root Mean Square metric.

A simpler version of the monomorphous TIE can be derived for weakly absorbing objects, when the approximation $I_0(x, y)/I_{in} = \exp b_0(x, y) \cong 1 + b_0(x, y)$, $b_0(x, y) = -2k\int_{-\infty}^{0} \beta(x, y, z)dz$, can be applied, i.e. when $|\ln[I_0(x, y)/I_{in}]| \ll 1$. In this case, eq.(11) can also be linearized:

$$Q_j(x,y) \cong 1 + b_j(x,y), \quad \text{where } b_j(x,y) = -2k \int \beta_j(x,y,z)dz, \quad j=1,2, \tag{24}$$

because $|b_j(x,y)| \leq |b_0(x,y)| << 1$. Substituting eq.(24) into eq.(18) and discarding the second-order terms (containing the products $b_i b_j$ or their derivatives), we obtain after simple algebraic transformations:

$$\begin{aligned} K_0 &\equiv I_0 / I_{in} - 1 = b_1 + b_2, \\ K_R &\equiv I_R / I_{in} - 1 = [1 - R\gamma_1/(2k)\nabla^2]b_1 + [1 - R\gamma_2/(2k)\nabla^2]b_2, \end{aligned} \tag{25}$$

where $K_0(x,y)$ and $K_R(x,y)$ are the experimentally measurable "contrast functions" in the object and image planes, respectively. As can be seen from eq.(25), in the case of weakly absorbing samples (note, that the phase shifts can in principle be large), the TIE becomes linear with respect to the absorption contributions of the two monomorphous components. Note that the equations for the zero-order (constant) Fourier components of $b_1$ and $b_2$ are under-determined, as for these components the first and the second line of eq.(25) give the same results, in agreement with the conservation of total intensity in the course of free-space propagation of light. Therefore, we can always assume without loss of generality, that, for example, the integral of $b_2$ over the image is equal to zero and the integral of $b_1$ is equal to the integral of $I_0 / I_{in} - 1$, i.e. to the total absorption in the sample.

The issue of numerical solution of eq.(25) is still not straightforward in general. Expressing $b_1$ from the first line of eq.(25) and substituting the result into the second line, leads to

$$K_R - [1 - R\gamma_1/(2k)\nabla^2]K_0 = R(\gamma_1 - \gamma_2)/(2k)\nabla^2 b_2. \tag{26}$$

Unfortunately, this equation is numerically unstable, similarly to eq.(19). However, compared to eq.(19), eq.(26) is easier to regularize (the fact that the integral of $b_2$ over the image is equal to zero can be used explicitly for that purpose, as shown below) and to solve. Collecting multiple images $I_R(x, y)$ at different propagation distances $R = R_k$ can also help in constructing a stable solution [24].

A regularized iterative approach, similar to the one given by eq.(22), can be devised by choosing an initial approximation $b_1^{(0)}(x, y) = f(x, y)$, where $f(x, y) = K_0(x, y)$ or $f(x, y) \equiv 0$ (or some a priori plausible distribution), and then iterating

$$b_2^{(k)} = [1 - R\gamma_2/(2k)\nabla^2]^{-1}\{K_R - [1 - R\gamma_1/(2k)\nabla^2]b_1^{(k-1)}\},$$
$$b_1^{(k)} = K_0 - b_2^{(k)}. \tag{27}$$

By substituting the second line of eq.(27) into the first one, it is possible to verify that this iterative scheme leads to the following series solution:

$$b_2 = K_0 + \sum_{n=0}^{\infty} \mathbf{A}^n \{\mathbf{B}K_R - K_0\}, \tag{28}$$

where $\mathbf{A} = [1 - R\gamma_1/(2k)\nabla^2][1 - R\gamma_2/(2k)\nabla^2]^{-1}$ and $\mathbf{B} = [1 - R\gamma_2/(2k)\nabla^2]^{-1}$. Operator $\mathbf{A}$ can be expressed as

$$\mathbf{A}f = \iint \exp[i2\pi(x\xi + y\eta)]\hat{f}(\xi,\eta)\frac{1 + \pi\lambda R\gamma_1(\xi^2 + \eta^2)}{1 + \pi\lambda R\gamma_2(\xi^2 + \eta^2)} d\xi d\eta. \tag{29}$$

Therefore, it is bounded in the space of square-integrable functions $L_2$ and its norm $\|\mathbf{A}\|_2$ does not exceed $[1+\pi\lambda R\gamma_1(\xi_{min}^2+\eta_{min}^2)]/[1+\pi\lambda R\gamma_2(\xi_{min}^2+\eta_{min}^2)]$. Let us consider for simplicity a square image $\Omega$ with the linear size $a$. Then the Fourier integral in eq.(29) can be replaced by the corresponding Fourier series. If we restrict the domain $D(\mathbf{A};\Omega)$ of functions, on which operator $\mathbf{A}$ acts, to the subspace $D_1(\mathbf{A};\Omega) = D(\mathbf{A};\Omega)\setminus\{\mathbf{1}\}$ equal to the orthogonal complement to constant functions, then we will have $\|\mathbf{A}\|_2 \leq [1+\pi\lambda R\gamma_1 a^{-2}]/[1+\pi\lambda R\gamma_2 a^{-2}] < 1$ on that subspace (as the lowest order of Fourier coefficients is now equal to one). This estimate guarantees the uniform and absolute convergence of the series in eq.(28) for any $f \in D_1(\mathbf{A};\Omega)$. We have explained above that it can be assumed without loss of generality that the zero-order Fourier coefficient of $b_2$ is equal to zero. As the matrix of the operator $\mathbf{A}$ is diagonal in the Fourier space representation (see eq.(29)), we can now find $b_2$ by solving eq.(28) on $D_1(\mathbf{A};\Omega)$. The series in eq.(28) then converge to

$$b_2 = 2k/[R(\gamma_1-\gamma_2)]\nabla^{-2}\{K_R - [1-R\gamma_1/(2k)\nabla^2]K_0 - c_0\}, \qquad (30)$$

where $c_0$ is a constant equal to the average of the function $K_R - [1-R\gamma_1/(2k)\nabla^2]K_0$ over the image $\Omega$. Once $b_2$ is found, $b_1$ can be easily found too from the first line of eq.(25). The phase function can then be obtained as $\varphi_0(x,y) = (\gamma_1/2)b_1(x,y) + (\gamma_2/2)b_2(x,y)$. Note that the solution given by eq.(30) corresponds to a direct regularization of eq.(26). However, as shown in the next section, a truncation of the series in eq.(28) (i.e. a finite number of iterations according to eq.(27)) may provide a more robust solution compared to eq.(30) in the presence of noise and experimental measurement errors (e.g. due to the changes in the incident illumination) in the input data (measured image intensities).

*5. Numerical tests*

In this section we present the results of a test of one of the phase-retrieval algorithms developed in the previous section. It is well known that the TIE-Hom phase retrieval method [8] is very stable and accurate when applied to monomorphous objects (corresponding to monomorphous complex amplitude distributions in the object plane). The "monomorphous decomposition" method developed in the present paper obviously reduces to TIE-Hom in the monomorphous case (here one can set $Q_1 \equiv 0$ or $Q_2 \equiv 0$). In order to investigate the most difficult case for the new method, we performed a test using a complex amplitude in the object plane that cannot be approximated by a monomorphous one. The relevant intensity and phase distributions are shown in Fig.2 (each image had 1024 x 1024 pixels). Obviously, here the logarithm of the intensity and the phase are not proportional to each other, so the complex wave amplitude is not monomorphous. The ratio $\gamma(x,y) = 2\varphi_0(x,y)/\ln I_0(x,y)$ varied between $\gamma_{min} = 12.1$ and $\gamma_{max} = 39.6$ in this example. For the reconstruction below, we chose $\gamma_1 = 10$ and $\gamma_2 = 50$, so that $\gamma_1 \leq \gamma_{min} < \gamma_{max} \leq \gamma_2$. We assigned the following physical parameters to the images: linear size was set to $a = 1$ cm, the range of intensity values was approximately (0.85, 0.9), the range of phase values was (-2.2, -1) and the wavelength was 1 Å (corresponding to hard X-rays). We then calculated an in-line free-space propagated image at the object-to-image distance $z = 10$ m by numerically evaluating the corresponding Fresnel integrals with the help of the well-tested X-TRACT software [37]. The corresponding image is shown in Fig.3. For this "ideal" (noise-free) image, the phase distribution in the object plane can be retrieved with high accuracy using a conventional TIE, eq.(17). We have verified this fact using an implementation of the general TIE solution available in X-TRACT. The relative $l_2$ - error between the original and the reconstructed phase distributions, calculated according to the usual formula,

$$d_2(\varphi_0^{rec}, \varphi_0^{true}) = \frac{\|\varphi_0^{rec} - \varphi_0^{true}\|_2}{\|\varphi_0^{true}\|_2} = \frac{\left(\sum_m \sum_n [\varphi_0^{rec}(m,n) - \varphi_0^{true}(m,n)]^2\right)^{1/2}}{\left(\sum_m \sum_n [\varphi_0^{true}(m,n)]^2\right)^{1/2}}, \quad (31)$$

was equal to 0.027 and the visual difference between the two images was imperceptible. We then added one percent (relative to the average image intensity) of Poisson noise to the intensity distributions in the object and image planes. The noisy intensity distribution in the object plane is shown in Fig.4. Even with this relatively small amount of noise, the performance of the conventional TIE phase retrieval (from images at two different propagation distances) deteriorated very significantly, see Fig.5. It is obvious from Fig.5a that there were strong errors in the low spatial frequencies (low-order aberrations). However, when the low-order components corresponding to the first 21 circular Zernike polynomials were subtracted from the reconstructed image, the higher order error terms became apparent (Fig.5b). The relative $l_2$ - error $d_2(\varphi_0^{rec}, \varphi_0^{true})$ between the original and the reconstructed phase distributions was equal to 3.19 here, i.e. it increased 118 times compared to the reconstruction from the noise-free images. Such a dramatic dependence of the reconstruction error on the noise in the input data is a well-known property of the TIE phase retrieval from images collected at different propagation distances (see e.g. [8, 9, 24]).

We then applied the iterative reconstructed algorithm defined by eq.(27) above to the intensity distributions in the object and image planes with 1% noise. The results are shown in Fig.6. Even though visually these reconstruction do not look much better (if at all) than the reconstructions in Fig.5 obtained using the conventional TIE, in fact the phase distributions in Fig.6 contain a much smaller amount of low-order aberrations and the overall the error $d_2(\varphi_0^{rec}, \varphi_0^{true})$ between the original and the reconstructed phase distributions was much smaller here: 0.883 and 0.642 for the distributions in Fig.6(a) and (b), respectively. Thus, the error in the phase reconstructed with eq.(27) was almost 5 times smaller compared to the reconstruction using the conventional TIE. We have also specifically compared the accuracy of the reconstruction of the low-order spatial frequencies of the phase distribution using the two methods. The sum of absolute errors in the first 21 Zernike coefficients between the original phase distribution and the one reconstructed using the conventional TIE was 4.144, while that error was equal to 1.081 and 0.991 in the images obtained using eq.(27) after 2 and 20 iterations, respectively.

The advantage of the method defined by eq.(27) over the phase retrieval using the conventional TIE was even more obvious in the case of geometrical misalignment between the images in the object and image planes. In order to simulate this problem, we shifted the image-plane intensity distribution by one pixel horizontally with respect to the object-plane intensity. This shift led to the $d_2(\varphi_0^{rec}, \varphi_0^{true})$ error in the conventional TIE reconstruction increasing by a further 50% from 3.19 (in the case of 1% noise and no shift) to 4.68 (in the case of 1% noise and 1 pixel shift). Remarkably, the accuracy of the reconstruction using eq.(27) changed by only a few percent as a result of this input data misalignment: $d_2(\varphi_0^{rec}, \varphi_0^{true})$ changed from 0.883 to 0.889 at 2 iterations, and from 0.642 to 0.718 at 20 iterations (see Table 1), the latter one being 6.5 times smaller than the corresponding error in the reconstruction using the conventional TIE.

Figure 7 shows the relative reconstruction error $d_2(\varphi_0^{rec}, \varphi_0^{true})$ as a function of the number of iterations (according to eq.(27)). One can see that the algorithm demonstrates a semi-convergent nature, as expected in the presence of noise and other inconsistencies in the input data. In fact, in this case, the series in eq.(28) still converges, but the limit no longer corresponds to the "true" (noise-free) phase distribution, because of the mathematical inconsistency of the input data due to the presence of noise and the geometrical misalignment. Therefore, in practice, when the ideal phase distribution is not known, the reconstruction can be stopped e.g. when the difference between two successive iterations becomes smaller than the noise level in the input data. This stopping criterion performed well in the case of the numerical example considered above.

## *6. Summary*

In this paper, we reviewed several types of objects (defined in terms of the spatial distribution of the complex refractive index) for which the quantitative analysis of in-line phase-contrast images and phase retrieval can be simplified. For monomorphous objects, in particular, the projected distribution of the complex refractive index can be uniquely reconstructed from measurements of in-line image intensity distribution in a single plane orthogonal to the optic axis in the near field [8]. We then demonstrated that an arbitrary pair of 2D distributions of phase and intensity in the object plane can always be represented as a linear combination of two monomorphous pairs of phase and intensity distributions. Such a decomposition of arbitrary complex wave amplitude (or, equivalently, of an arbitrary 3D distribution of the complex refractive index in the case of CT) can be used as a basis for development of "stabilized" versions of phase retrieval algorithms based on the TIE. In our numerical tests, using a proposed iterative algorithm based on the monomorpous decomposition, the reconstruction of the phase distribution from in-line intensities in the object and image planes has demonstrated an improved stability in the case of the input data containing simulated photon noise and geometrical misalignment. The reconstruction was also quite stable as a function of the number of iterations. As the proposed method appears capable of providing better accuracy in phase retrieval compared to the conventional algorithms, we believe that it can be useful in quantitative 2D phase-contrast imaging and in phase-contrast tomography.

Table 1. Relative $l_2$ - errors between the original and the reconstructed phase distributions obtained using eq.(27) with different number of iterations. The input data contained 1% Poisson noise (column 2) and an additional 1 pixel horizontal shift of the propagated image (column 3).

| Number of iterations, eq.(27) | $d_2(\varphi_0^{rec}, \varphi_0^{true})$ error (1% noise) | $d_2(\varphi_0^{rec}, \varphi_0^{true})$ error (1% noise and 1 pixel shift) |
|---|---|---|
| 2 | 0.883 | 0.889 |
| 4 | 0.814 | 0.826 |
| 6 | 0.770 | 0.789 |
| 10 | 0.713 | 0.747 |
| 20 | 0.642 | 0.718 |
| 50 | 0.576 | 0.792 |
| 100 | 0579 | 1.010 |

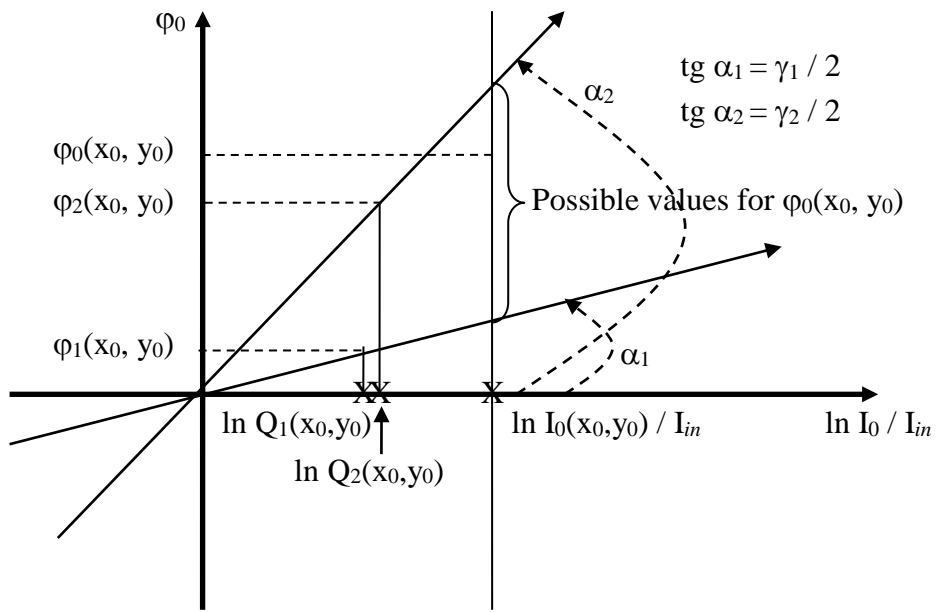

Fig. 1. Illustration of the constraints imposed by *a priori* knowledge of coefficients $\gamma_1$ and $\gamma_2$ on the range of possible phase values corresponding to a given intensity value.

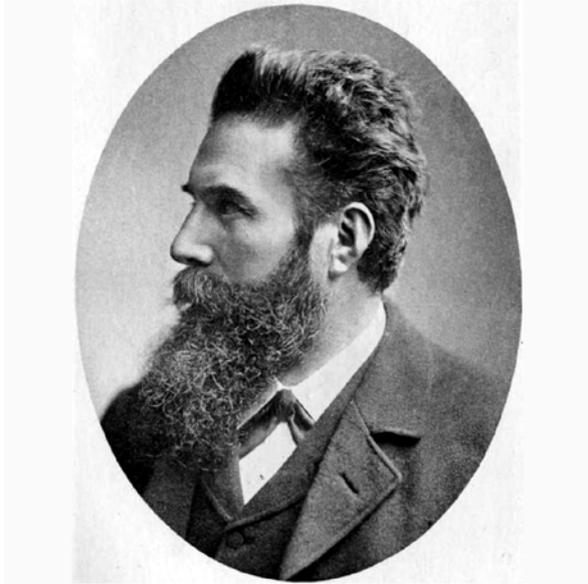 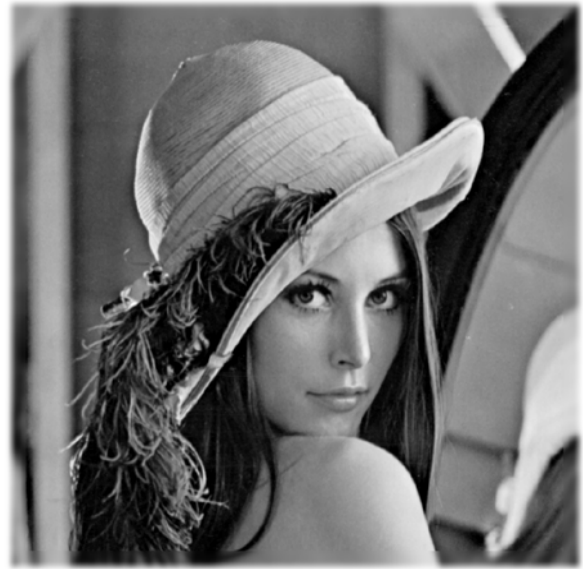

(a) (b)

Fig.2a. Initial intensity, (a), and phase, (b), distributions in the object plane.

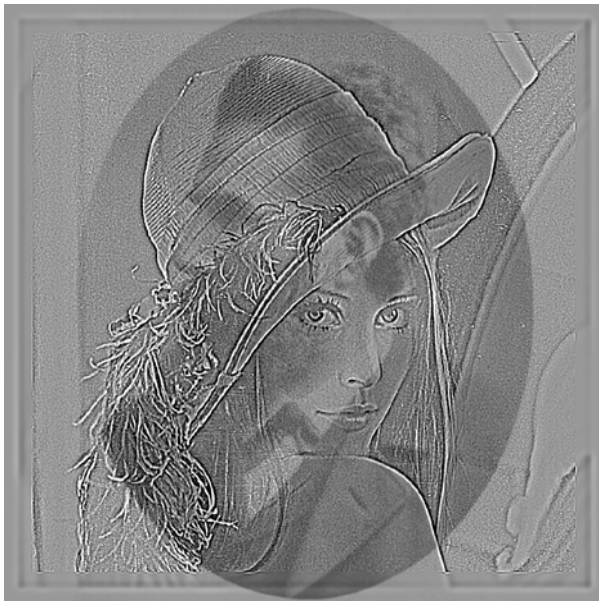 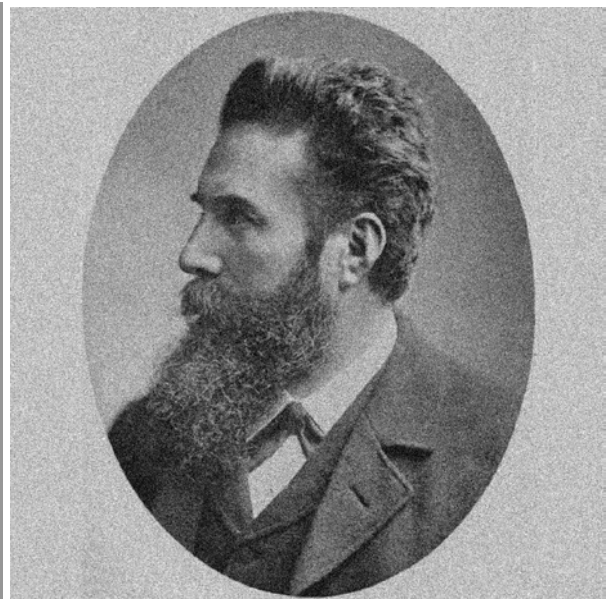

Fig.3. In-line image intensity distribution.    Fig.4. Object plane intensity with 1% noise.

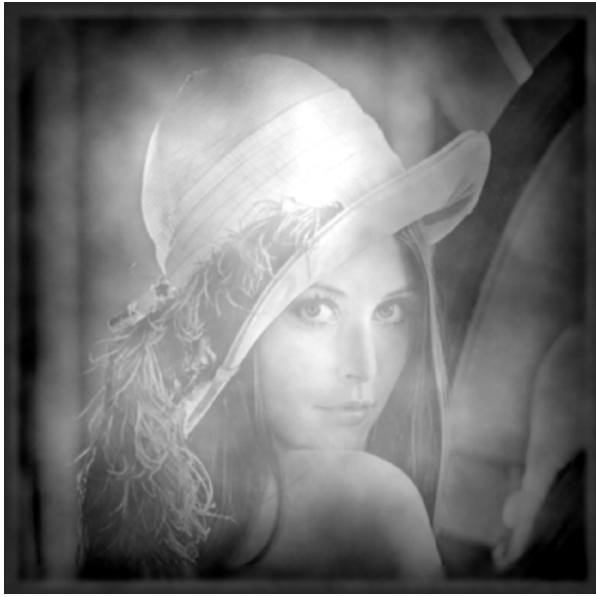 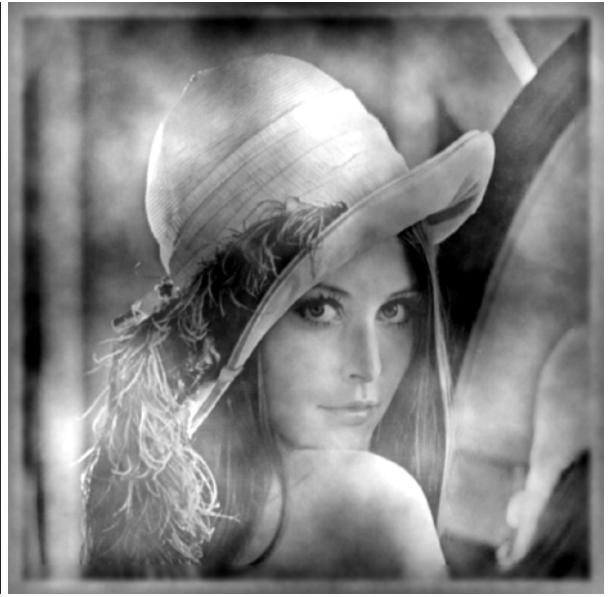

(a) (b)

Fig.5. Phase distribution in the image plane reconstructed using conventional TIE, eq.(17), from two images with 1% noise, (a); the same reconstructed distribution with the first 21 low-order Zernike components subtracted, (b).

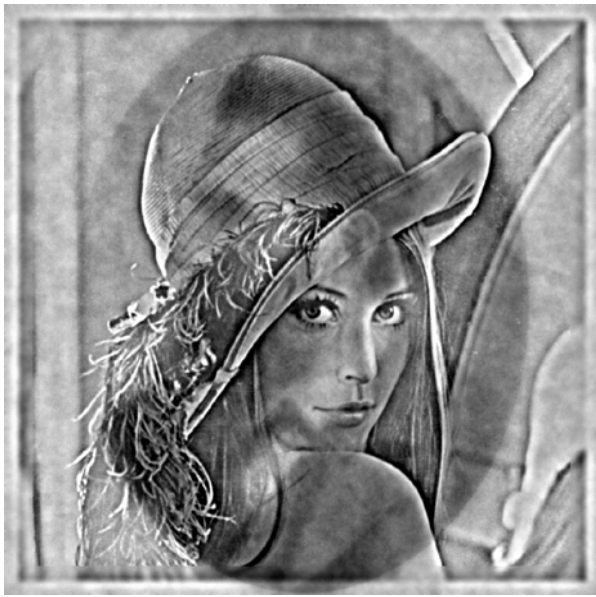 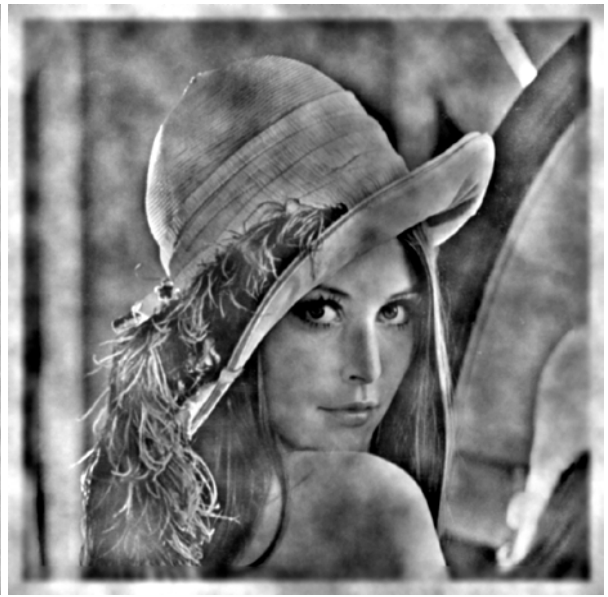

(a) (b)

Fig.6. Phase distribution in the image plane reconstructed according to eq.(27) from two images with 1% noise, after 2 iterations (a), and after 20 iterations, (b).

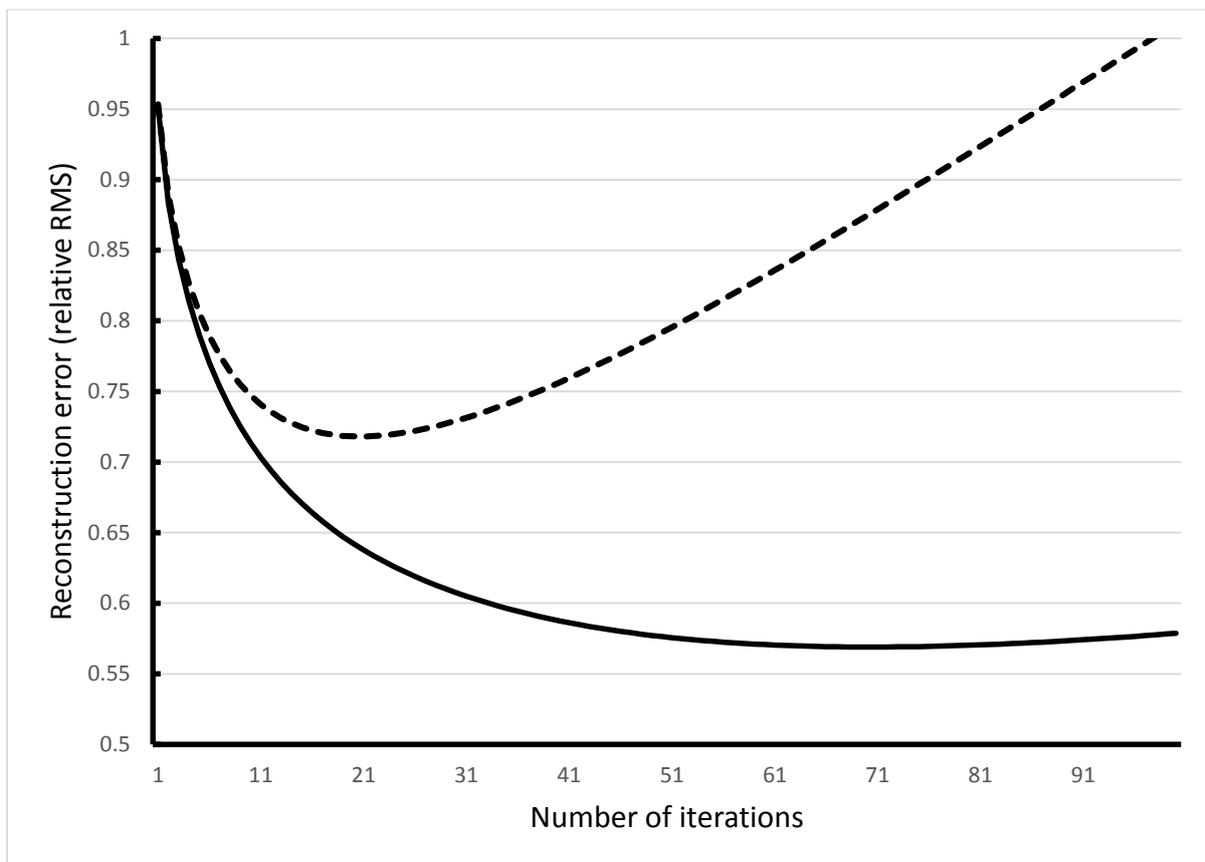

Fig. 7. Reconstruction error as a function of the number of iterations (solid line - 1 % noise in input data, dashed line - 1 % noise and 1 pixel horizontal shift).